\documentclass[12pt,preprint,natbib]{aastex}

\begin{document}
\title{Direct Imaging Confirmation and Characterization of a Dust-Enshrouded Candidate Exoplanet Orbiting Fomalhaut}
\author{Thayne Currie\altaffilmark{1,2}, 
John Debes\altaffilmark{3},
Timothy J. Rodigas\altaffilmark{4},
Adam Burrows\altaffilmark{5},
Yoichi Itoh\altaffilmark{6}, 
Misato Fukagawa\altaffilmark{7}, 
Scott J. Kenyon\altaffilmark{8}, 
Marc Kuchner\altaffilmark{2},
Soko Matsumura\altaffilmark{9}} 
\altaffiltext{1}{University of Toronto}
\altaffiltext{2}{NASA-Goddard Space Flight Center}
\altaffiltext{3}{Space Telescope Science Institute}
\altaffiltext{4}{Steward Observatory, University of Arizona}
\altaffiltext{5}{Department of Astrophysical Sciences, Princeton University}
\altaffiltext{6}{Nishi-Harima Observatory, University of Hyogo}
\altaffiltext{7}{Osaka University}
\altaffiltext{8}{Smithsonian Astrophysical Observatory}
\altaffiltext{9}{Department of Astronomy, University of Maryland-College Park}
\email{currie@astro.utoronto.ca}
\begin{abstract}
We present Subaru/IRCS J band data for Fomalhaut and a (re)reduction of archival 
 2004--2006 HST/ACS data first presented by 
Kalas et al. (2008).  We confirm the existence of a candidate 
exoplanet, Fomalhaut b, in both the 2004 and 2006 F606W data sets at a 
high signal-to-noise.  Additionally, we confirm the detection at F814W and present a new detection in F435W.  
  Fomalhaut b's space motion may be consistent with it 
being in an apsidally-aligned,  non debris ring-crossing orbit, although new astrometry 
is required for firmer conclusions.  
We cannot confirm that Fomalhaut b exhibits 0.7-0.8 mag variability cited as evidence for planet accretion or 
a semi-transient dust cloud.  The new, combined optical SED and IR upper limits confirm that emission identifying Fomalhaut b originates 
from starlight scattered by small dust, but this dust is most likely associated with a massive body.
The Subaru and IRAC/4.5 $\mu m$ upper limits imply $M$ $<$ 2 $M_{J}$, 
still consistent with the range of Fomalhaut b masses needed to sculpt the disk.
Fomalhaut b is very plausibly ``a planet identified from direct imaging" even if current images of it do not, strictly 
speaking, show thermal emission from a directly imaged planet.
\end{abstract}
\section{Introduction}
In November 2008, two teams announced what were described as the first directly imaged extrasolar planets 
around the nearby young stars HR 8799 and Fomalhaut \citep{Marois2008,Kalas2008}.  
Numerous studies recovered the three HR 8799 planets in 
archival data \citep[][]{Lafreniere2009,Fukagawa2009,Metchev2009,Currie2012a}; recent 
ones identify a fourth planet (HR 8799e, Marois et al. 2010; Currie et al. 2011a) 
with a similar spectrum \citep{Skemer2012,Esposito2012}.
  Detailed dynamical and atmospheric modeling 
confirms that HR 8799 bcde are bona fide planets, not brown dwarfs 
\citep[$M_{bcde}$ $\le$ 9 $M_{J}$,][]{Marois2011,Currie2011a,Sudol2012}.  

In contrast, refereed follow-up studies have yet to confirm 
Fomalhaut b, casting serious doubt on its status.
As noted by \citeauthor{Kalas2008}, Fomalhaut b's spectrum looks little like that expected 
for a substellar object, as it is detected in the optical 
but not in the infrared.
Deeper IR data do not recover Fomalhaut b \citep[][]{Marengo2009,Janson2012}.  
 \citeauthor{Kalas2008} describe Fomalhaut b's emission as being 
dominated by scattered dust emission surrounding a planet, with a thermal component 
at 0.8 $\mu m$ and accretion-driven variability at 0.6 $\mu m$.
 However, \citet{Janson2012} argue, based on their detection limits, that essentially 'none' 
of the optical emission originates from a planet, the 0.6 $\mu m$ variability instead identifies 
a semi-transient dust cloud and thus Fomalhaut b is not a directly imaged planet.
While Fomalhaut b seems to be a natural explanation for the well-confined 
debris ring \citep{Kalas2005}, shepherding planets unassociated 
with Fomalhaut b may provide an alternate explanation \citep{Boley2012}.  

These doubts about Fomalhaut b's 
status call for new data to further constrain its IR emission and a reanalysis of the
HST data from which Fomalhaut b was originally identified.
Thus, in this Letter, we present new Subaru/IRCS J band imaging of Fomalhaut and 
a rereduction of \citeauthor{Kalas2008}'s HST/ACS data.
  We use these data to constrain the near-IR emission 
from any planet corresponding to Fomalhaut b and revisit key results of 
\citet{Kalas2008}: 1) Fomalhaut b's existence, 2) its spectrum, 3)
 0.6 $\mu m$ variability, and 4) its orbit.

\section{Observations and Data Reduction}

 \textbf{Subaru/IRCS Data} -- 
We imaged Fomalhaut on 2009 Aug 16 in the Mauna Kea $J$ band filter behind 
the 0\farcs{}8 coronographic spot 
with the Infrared Camera and Spectrograph \citep[IRCS,][]{Tokunaga1998} and the AO-188 adaptive 
optics system (full-width half-maximum (FWHM) = 0\farcs{}06; 20.53 mas pixel$^{-1}$) 
in observing blocks bracketing HR 8799 observations \citep{Currie2011a} 
in \textit{angular differential imaging} (ADI) mode \citep{Marois2006}.  
 Between the two observing blocks, we rotated the instrument by $\sim$ 70$^{o}$ to 
keep the region covering Fomalhaut b visible on the detector ($\delta$PA $\sim$ 94$^{o}$).
 Basic image processing follows steps listed in \citet{Currie2010,Currie2011a,Currie2011b}.

\textbf{HST/ACS Data and Basic Processing} -- We downloaded HST/ACS data for Fomalhaut taken in the F435W, F606W, and F814W filters taken 
in 2004 and 2006 (Program IDs 10390 and 10598) from the MAST archive and processed with the OPUS pipeline.

The distortion-corrected ACS data are pocketed by cosmic-ray hits.
We flag most of these cosmic rays in individual subimages by identifying 10-$\sigma$ 
outliers in concentric annuli roughly 
centered on the star in a median-filtered image with a moving-box length of 10 pixels 
We register images by subtracting the first image in the sequence by a 180$^{o}$ rotation of itself,
 identifying the centroid position 
that minimizes the residuals, and then cross-correlating this image with the others to derive 
relative offsets.  We reform the subimages together into datacubes,
mean-combining them with $n$-$\sigma$ outlier clipping to mitigate other cosmic-rays 
and hot pixels, iteratively choosing the exact settings 
to maximize the signal-to-noise of background objects (derived at the end of our image processing: 
next subject heading). 

\textbf{PSF Subtraction} -- We explored a range of methods to subtract the stellar 
point-spread function (PSF), including 
reference PSF subtraction, roll subtraction/ADI-based PSF subtraction, LOCI \citep[``Locally Optimized 
Combination of Images"][]{Lafreniere2007}, and 
adaptive LOCI or A-LOCI \citep[][T. Currie, 2012 in prep.]{Currie2012a}.  
For the Subaru data, we obtain the best sensitivity using 
LOCI (A-LOCI yielded no improvement).  For the F435W and F814W data, the 
small number of images or random noise sources reduce (A)LOCI's effectiveness.  
We obtain the smallest subtraction residuals by performing roll subtraction 
separately for each integration time and averaging the results.

For the F606W data, A-LOCI provides the best results.  The \citeauthor{Currie2012a} version filters 
 images by their degree of speckle correlation and determines the combination of algorithm parameters that 
maximize the SNR of point sources.  Here, we add two components.
First,  we add a ``moving pixel mask", where the algorithm is prevented from using, in the PSF construction, 
pixels lying within the annular region we want to subtract (the 'subtraction zone', see 
Marois et al. 2010b; Soummer et al. 2011).  Second, we also include frames from the reference star Vega in 
our A-LOCI PSF construction \citep[a ``PSF reference libary";][]{Soummer2011}. 

\section{Results}
\subsection{Detections, Astrometry and Photometry}
\begin{figure}
\centering
\includegraphics[scale=.4011,trim = 15mm 0mm 20mm 2mm,clip]{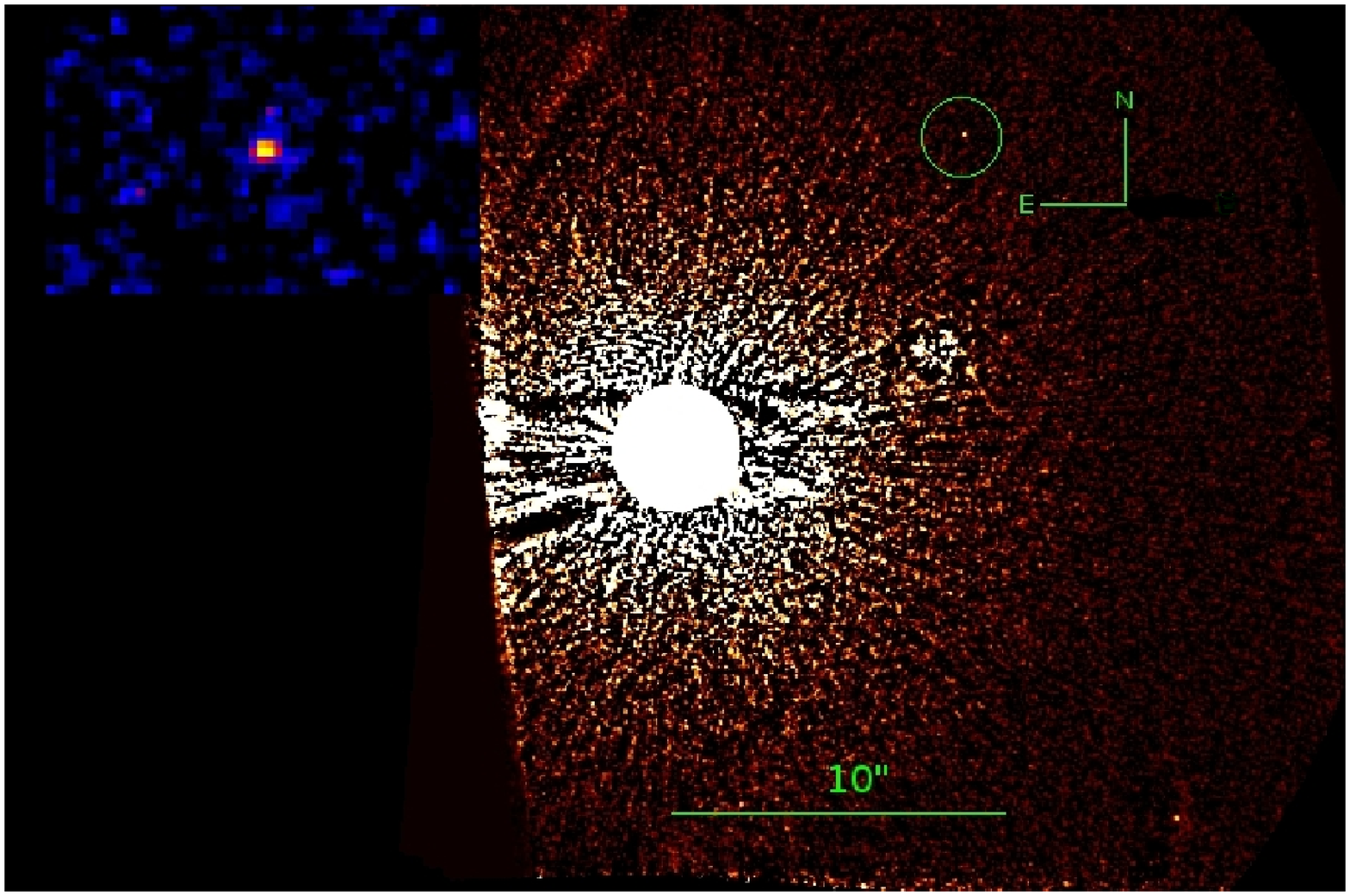}
\includegraphics[scale=.399,trim = 0mm 0mm 50mm 0mm,clip]{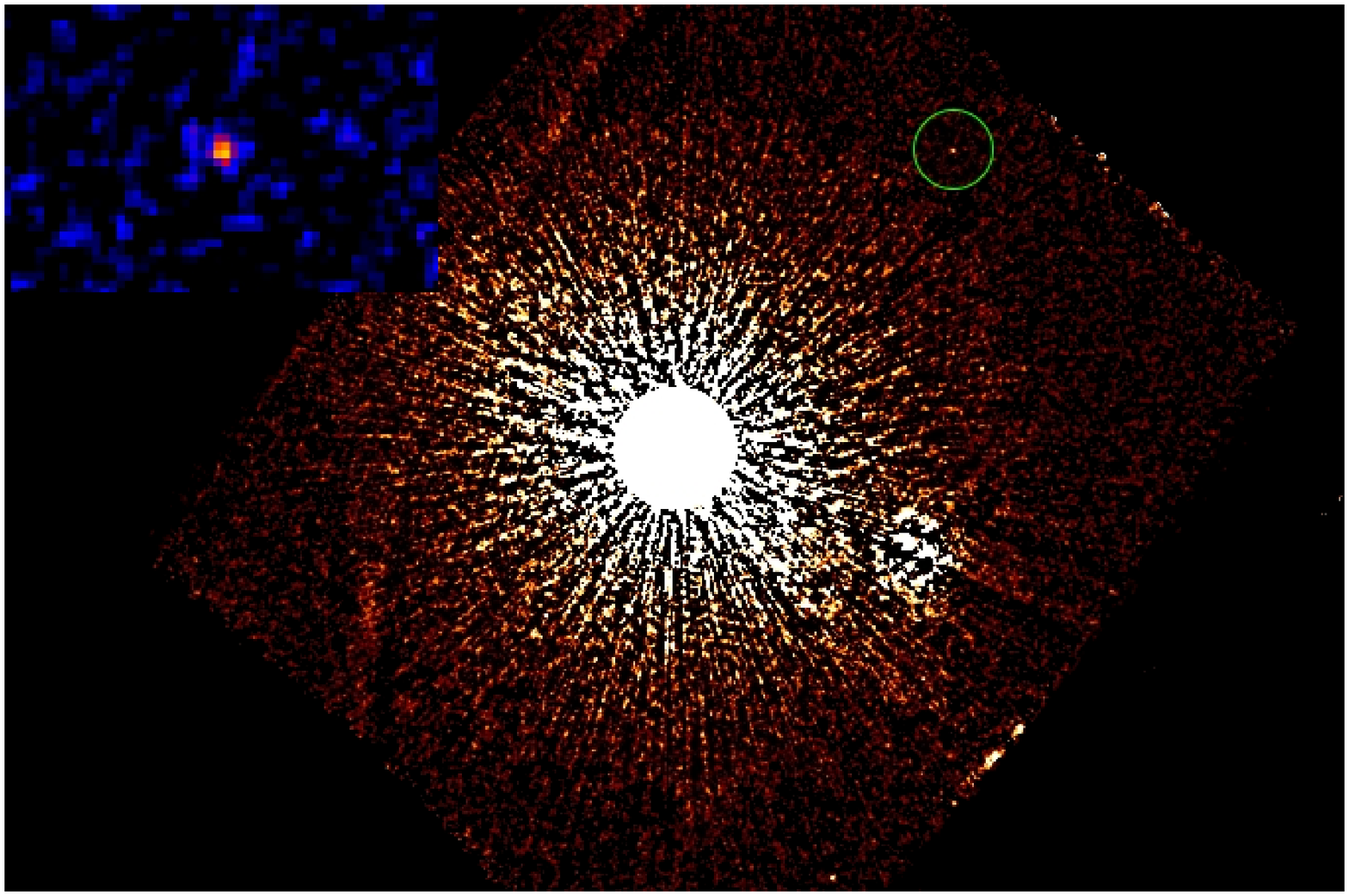}
\\
\includegraphics[scale=0.55,trim=0mm 0mm 0mm 0mm,clip]{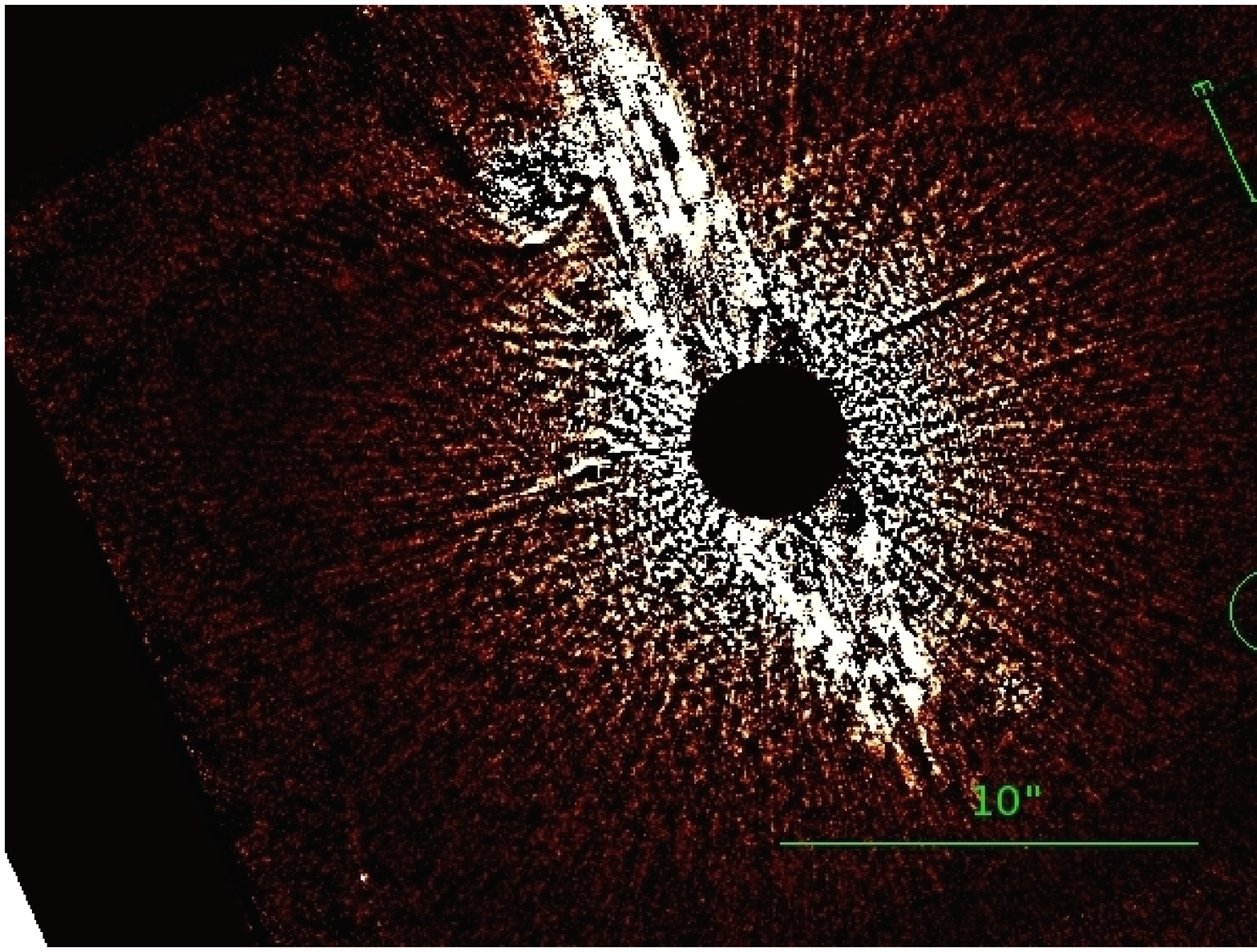}
\caption{ \textbf{\textit{A-LOCI}} processed images of Fomalhaut from 2006 (top-left; 3\farcs0 coronagraph) and 2004 
(top-right; 1\farcs8 coronagraph) F606W data.  The insets in the upper lefthand corner 
are centered on Fomalhaut b and have a box length of $\sim$ 0\farcs75 by 0\farcs5.
Image combining the 1\farcs8 and 3\farcs0 2006 data rotated by 66$^{o}$ and enlarged with the inner 
$r$ $<$ 1\farcs75 region masked.}
\label{f606wim}
\end{figure}

As shown in Figure \ref{f606wim}, 
we easily recover Fomalhaut b in all F606W data sets.  With the noise defined in concentric annuli, 
the signal-to-noise ratios (SNR) are 
SNR = 16 for the 3\farcs{}0 2006 data (top-left panel), 9 for the 
1\farcs{}8 2004 data (top-right panel), and $\sim$ 8 for the 2006 1\farcs{}8 data (not shown);
 the corresponding A-LOCI parameters are $\delta$ = [1,1,1], $N_{A}$ = [125, 150, 125], $g$ =[1,1,1], 
$dr$ = [10,5,5], and $r_{corr}$ = [0.9,0.5,0.7].
Defining the significance of our detection within a 5--10 pixel annulus surrounding Fomalhaut b,  we get SNR $\sim$ 
10.2, 14.1, and 16.4 for the 2006 1\farcs{}8 data, 3\farcs{}0 data, and the combined data (bottom panel) and SNR $\sim$ 12.7 for the 
2004 data.  These estimates are about a factor of 2 higher than the best SNR estimates from \citeauthor{Kalas2008} (SNR $\sim$ 8; 
P. Kalas, pvt. comm.).   
The images include a background star ($r, PA$ (2006) = 14\farcs2, 203$^{\circ}$)
 and two background galaxies ($r, PA$ (2006) = 10\farcs8, 135$^{\circ}$; 
$r, PA$ (2006) = 18\farcs7, 234$^{\circ}$).  We do not conclusively identify 
any other candidate companions at 5-$\sigma$, including at the location of 
\citet{Janson2012}'s candidate nor within the debris ring.

\begin{figure}
\centering
\epsscale{1}
\includegraphics[scale=0.55,trim= 35mm 15mm 40mm 15mm,clip]{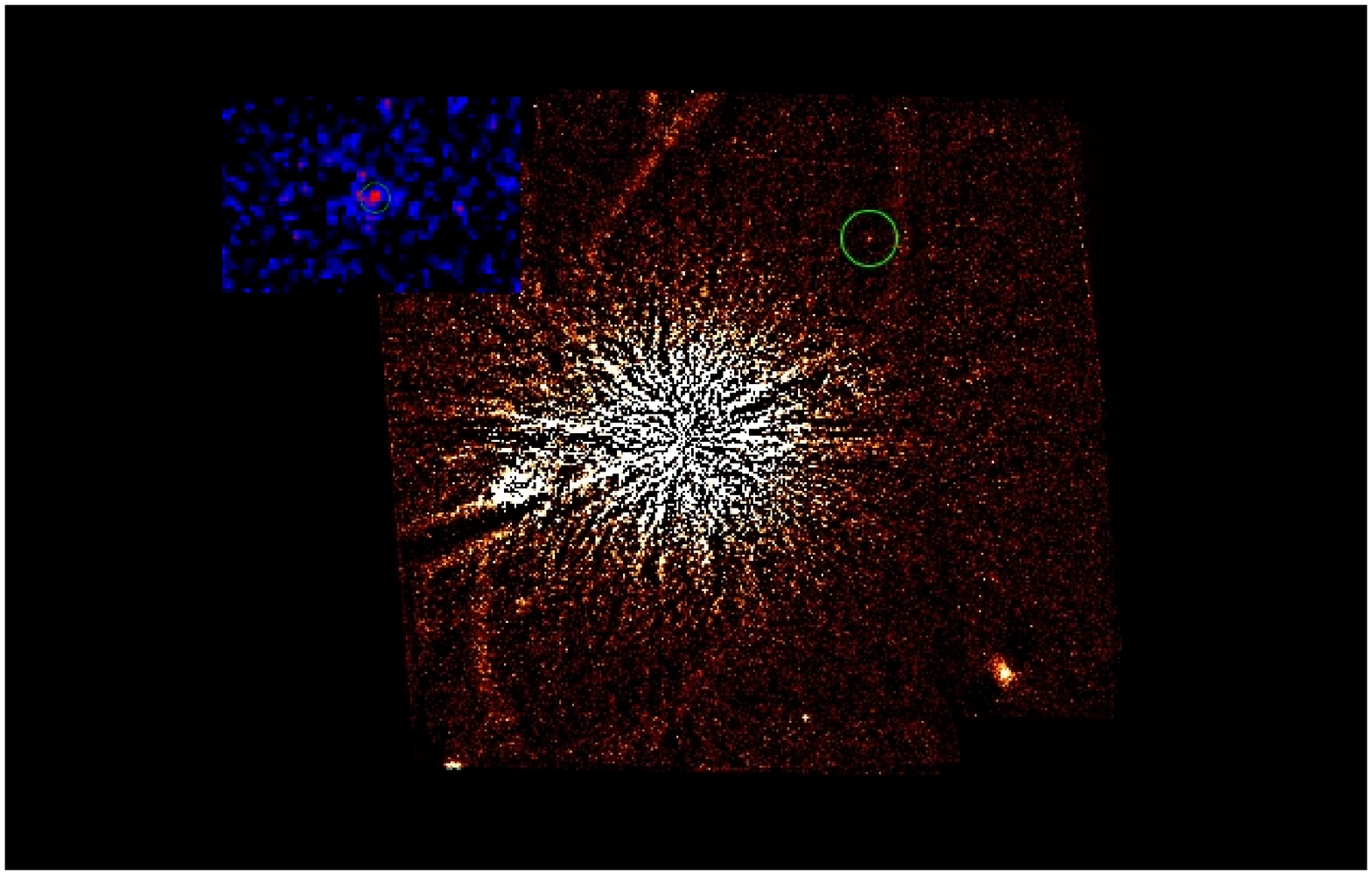} 
\includegraphics[scale=0.55,trim= 0mm 3mm 10mm 0mm,clip]{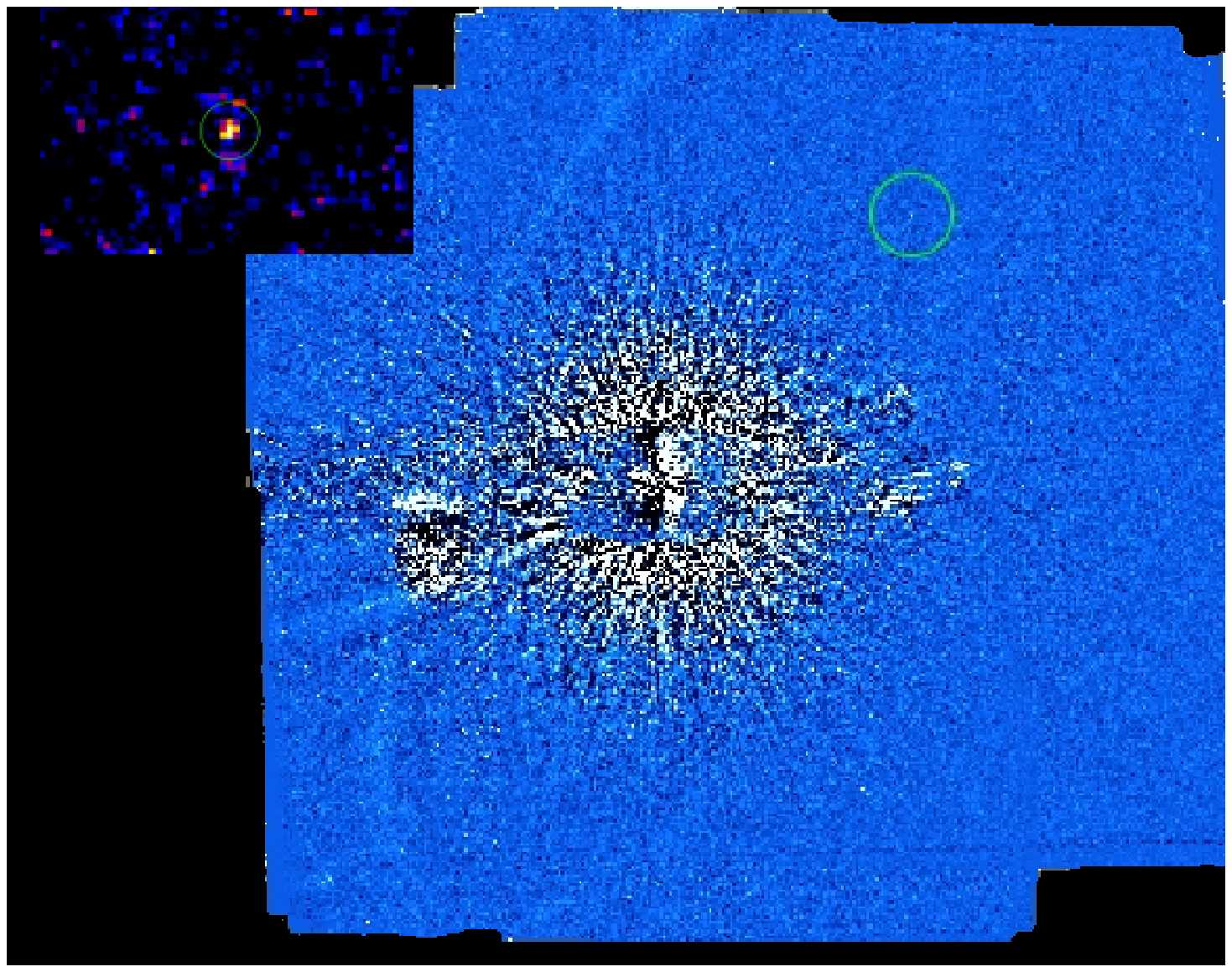} 
\includegraphics[scale=0.6,trim=30mm 0mm 30mm 60mm,clip]{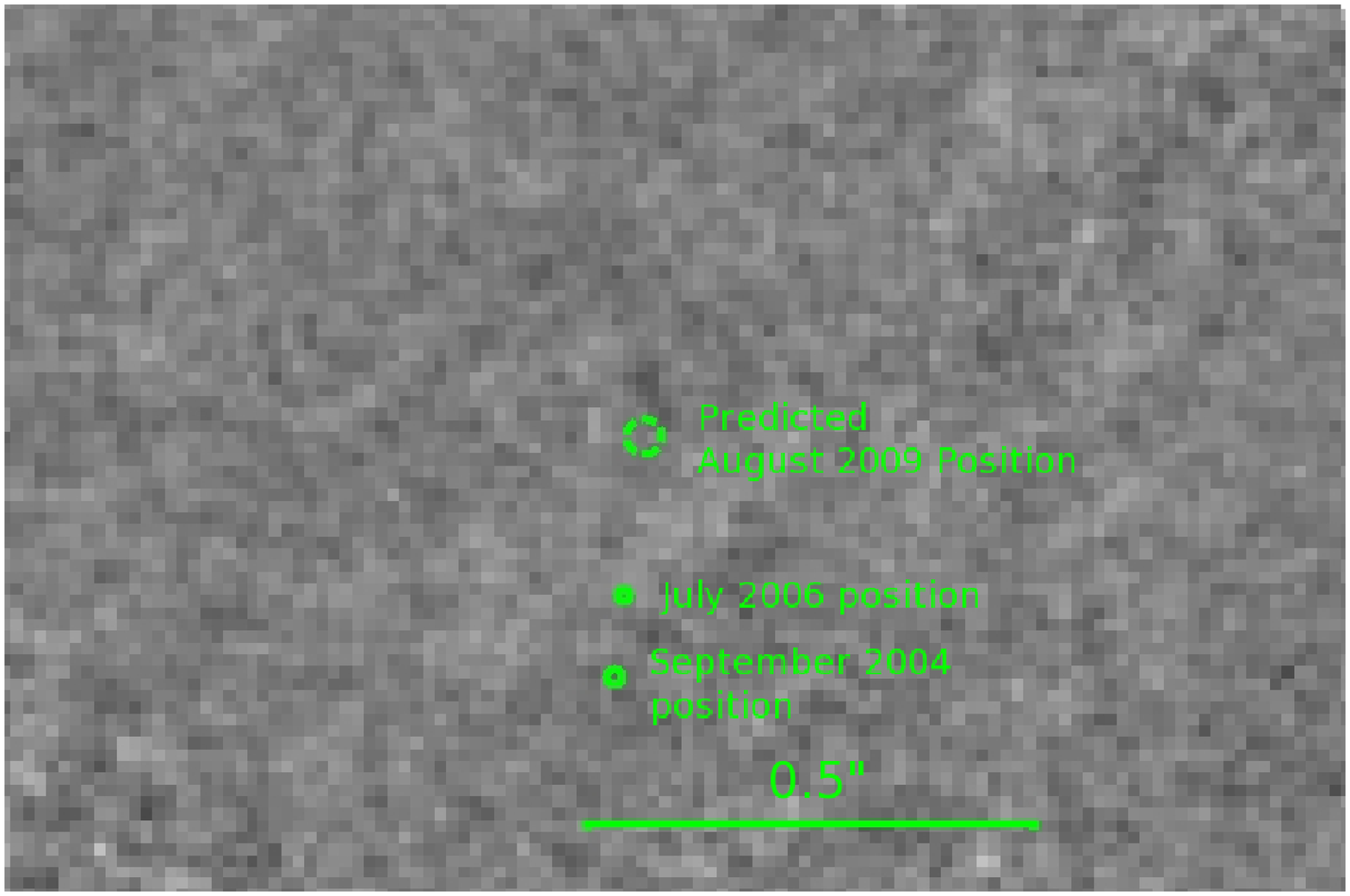}
\caption{\textbf{\textit{Roll-subtracted}} images of Fomalhaut in the F814W (top-left) and F435W (top-right) filters
from 2006 HST data and \textit{\textbf{LOCI-processed}} $J$ band/1.25 $\mu m$ from 2009 
Subaru/IRCS data 
(bottom).  The image scale for the F814 and F435 data is the same as in Figure 1; the IRCS image 
focuses on a small, $\sim$ 1\farcs0 by 0\farcs75 region centered on the measured and predicted positions 
for Fomalhaut b, with the 2009 (2004, 2006) circle size equal to the image FWHM (HST astrometric errors).
}
\label{otherwvlh}
\end{figure}

We also recover the F814W detection reported by \citeauthor{Kalas2008} and
 report a new F435W detection of Fomalhaut b (SNR $\sim$ 6) (Figure \ref{otherwvlh}).  
However, we do not detect Fomalhaut b in the Subaru/J band data (Figure \ref{otherwvlh}).

To measure Fomalhaut b's brightness in the HST data and to quantify Subaru upper limits, we 
use aperture photometry and determine/correct for throughput losses as in \citet{Lafreniere2007}.  
Fomalhaut b is nearly unattenuated in A-LOCI reductions, a fact confirmed by comparing the F606W brightness of Fomalhaut b 
with photometry using reference PSF subtraction.  For the 2006 F606W data, we derive a brightness 
of $m$ = 24.97 $\pm$ 0.09.  
In the 2004 F606W data, we derive $m$ = 24.92 $\pm$ 0.10.  Fomalhaut b has 
not measurably varied in brightness between the two epochs, a result that appears insensitive 
to our choice of reduction method. Photometry at F435W and F814W imply that Fomalhaut b is near 
zero color ($m_{F435W}$ = 25.22 $\pm$ 0.18, $m_{F814W}$ = 24.91 $\pm$ 0.20).
Our 5-$\sigma$ Subaru/J band limits are $m$ $>$ 22.22.

We determine Fomalhaut b's position from the highest SNR data 
at each epoch (F606W), fitting a Gaussian profile (FWHM = 2.7 pixels).  The astrometric uncertainties 
consider the intrinsic SNR, our precision in deriving 
absolute astrometric calibration, and differences in positions derived from the full range of 
PSF subtraction methods (reference PSF subtraction, LOCI, A-LOCI).  
Fomalhaut b's position changes in the 2006 (2004) data by $\sigma$[E,N] $\sim$ [10, 10] mas ([15, 5] mas) 
depending on exactly how we do the PSF subtraction.  The SNR for Fomalhaut b and background objects 
is insensitive ($\delta$SNR $<$ 5--10\%) to centroid offsets $<$ 13 mas (0.5 pixels) from the position that 
minimizes the subtraction residuals (see \S 2, paragraph 3).  

Our analyses yield a 2006 position of [E,N]\arcsec{} = [-8.615, 9.352]\arcsec{} 
$\pm$ [0.016, 0.016]\arcsec{} and 2004 position of [-8.598, 9.209]\arcsec{} $\pm$ [0.020, 0.014].
\citeauthor{Kalas2008}'s astrometry shows good agreement with ours in the 
2004 $North$ coordinate but otherwise differs by $\sim$ 0.5--1 pixels.  Their astrometry also implies 
a larger space motion of $\approx$ 0\farcs{}18 vs. our $\approx$ 0\farcs{}14.
\subsection{Photometric and Astrometric Modeling}
\subsubsection{SED Modeling: Planet Mass Upper Limits}
To interpret the infrared non-detections, 
we use the planet atmosphere models in
\citet{Spiegel2012} for 1--10 $M_{J}$ companions and \citet{Baraffe2003} for 0.5--1 $M_{J}$ companions.
Our comparisons help to determine whether any planet masses are consistent with 
the combined upper limits and with the range of masses needed for Fomalhaut to sculpt the debris ring.
Here, we consider an age range of 100--500 Myr, encompassing 
age estimates from individual indicators in \citet{Mamajek2012}, where the 
best-estimated age is $\sim$ 450 Myr.

\begin{figure}
\centering
\epsscale{1}
\plottwo{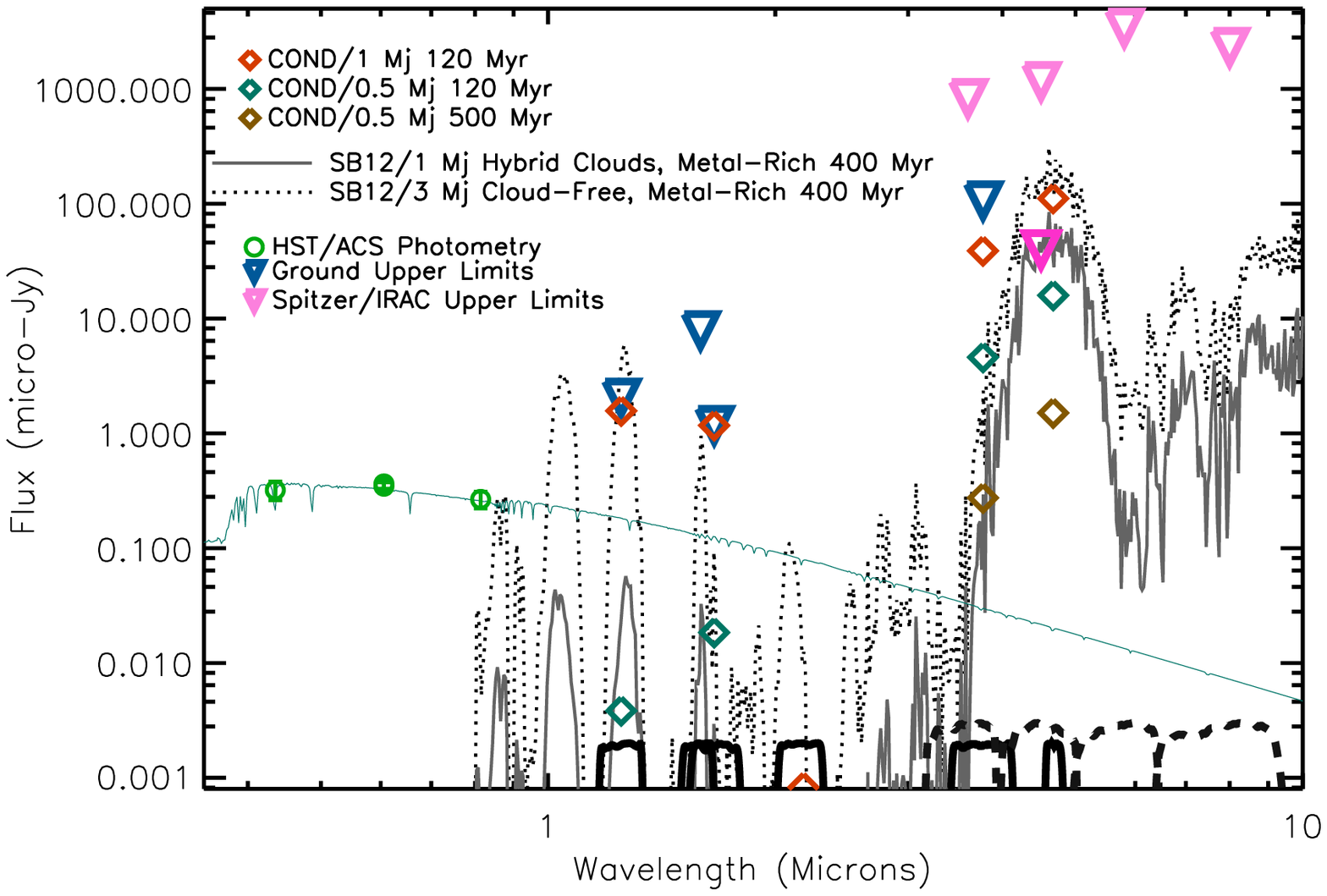}{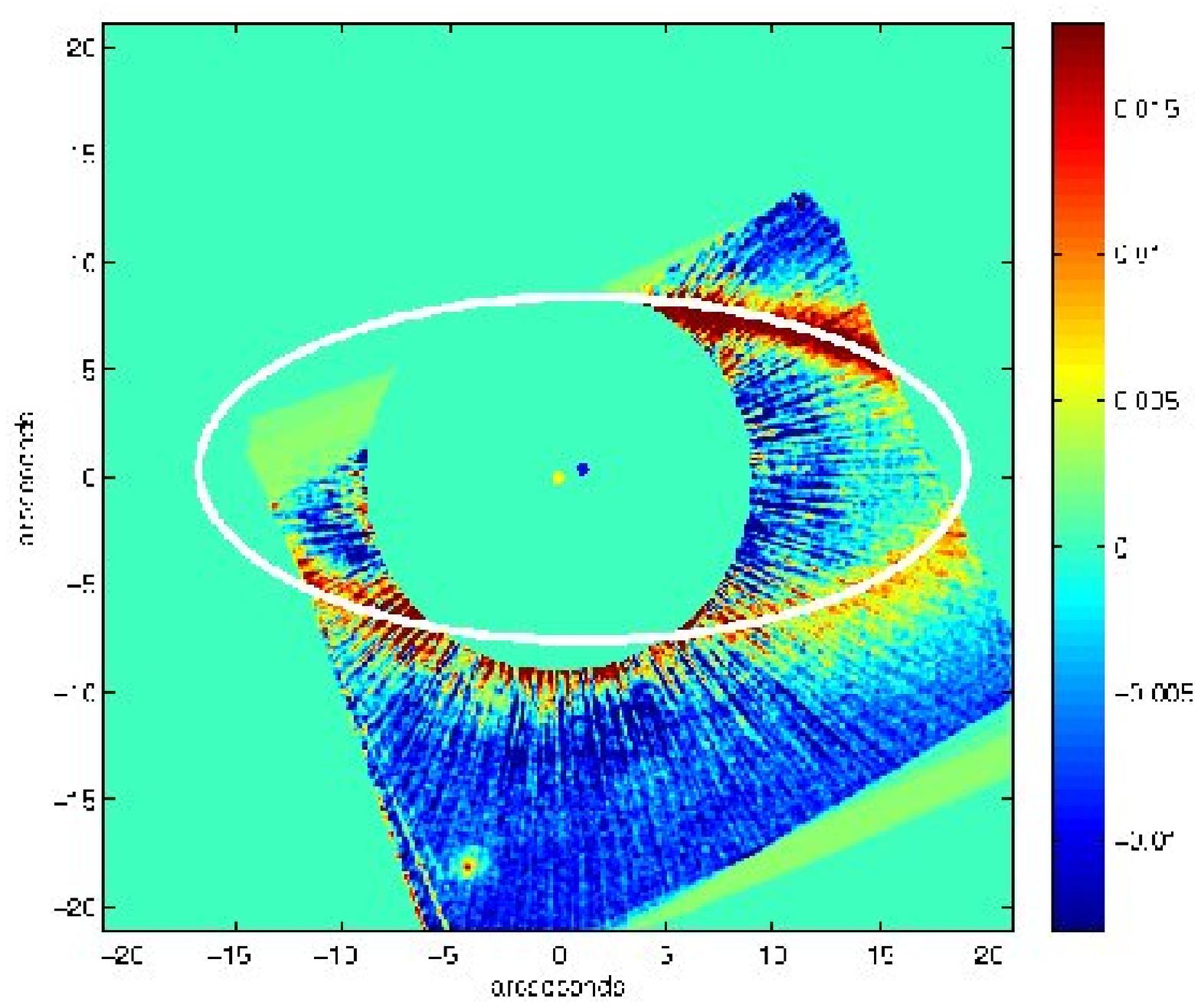}
\caption{(Left) Fomalhaut b photometry compared to model spectra from \citet{Spiegel2012} and 
photometry predictions from \citet{Baraffe2003} 
for 0.5--1 $M_{J}$ planets at 120-500 Myr and fit with a scaled mid-A star photosphere (blue-green spectrum).  The 
 thick black lines and dotted black lines depict the transmission functions for the ground-based and Spitzer 
passbands, respectively. (Right) Ellipse fit to the debris ring from reference PSF-subtracted 
3\farcs0 F606W Fomalhaut data from Method 1 (units of counts/s).  The yellow (blue) dot identifies the star (disk center).  
Like Figure 1 (bottom panel), the image is rotated by 66$^{o}$.}
\label{seds}
\end{figure}

Figure \ref{seds} compares several classes of models to the
 optical detections and the IR upper limits. The optical SED matches
the spectrum of light from Fomalhaut (blue-green line) scattered by small dust.  
Optical photospheric predictions for the planet models are 
more than 5 magnitudes fainter than the 0.4--0.6 $\mu m$ emission.   
Although the optical 
detections lie well above model predictions for gas giant planets, 
\citet{Janson2012}'s the 5-$\sigma$ IRAC upper limit 
  restricts Fomalhaut b's mass to 
less than 2 $M_{J}$ for even the oldest ages.  The $J$-band data also restrict Fomalhaut b's mass to be 
$<$ 2 $M_{J}$ for younger ages, though useful limits at 400--500 Myr are entirely 
due to the IRAC data.

However, the ground-based/Spitzer IR limits are still bright enough to be 
consistent with Fomalhaut b being a gas giant planet that sculpts the debris ring.  The 5-$\sigma$ IRAC upper limits 
are comparable to the predicted brightness of a 400 Myr-old, 1 $M_{J}$ planet from 
both the \citet{Spiegel2012} and \citet{Baraffe2003} models.
The \citet{Janson2012} upper limits 
are still $\sim$ 3 (30) times brighter than a 0.5 $M_{J}$ planet at (120) 500 Myr, which 
is massive enough to sculpt the debris ring:
 in fact, this low mass is favored by detailed dynamical simulations 
\citep{Chiang2009}.  

\subsubsection{Nature of the Fomalhaut b Optical Emission}
Following \citet{Kalas2008}, a dust cloud producing Fomalhaut b's optical emission ($m_{F606W, avg.}$ = 24.95) 
at 119 AU requires a minimum cross-sectional area of $\sigma_{p}$ $\approx$ 1.3$\times$10$^{18}$/$Q_{s}$ 
$\approx$ 1.3$\times$10$^{18}$ 
$m^{2}$.  For spherically distributed dust, this implies a dust cloud radius of 
$r_{c}$ $\gtrsim$ 0.0043 AU. 
Because Fomalhaut b is unresolved, the projected emitting radius of the cloud at F606W must be 
$a$ $\lesssim$ 0.5 FWHM or $\approx$ 0.26 AU.
The timescale for orbital shear to broaden a dust cloud of initial radius $r_{c}$ 
into a region of radius $a$ with orbital period $P$ is $t_{m}$ $\approx$ ($a$/$r_{c}$)P
\citep{KenyonBromley2005}. 

From this Keplerian shear alone, an unbound dust cloud at the separation of Fomalhaut b 
would then smooth out into a resolvable ($a$ $>$ 0.26 AU) clump in 56,000 yr and a circumstellar 
ring in $\approx$ 25 Myr ($\approx$ 1/18th of Fomalhaut's age).
Other dust grain removal mechanisms (radiation pressure, Poynting-Robertson drag) operate on similar/shorter timescales.
\textbf{\textit{Unless we are fortuitously (and implausibly) identifying a very recent collision}} 
in a region with $>$ 10$\times$ lower optical depth (and thus far lower collision frequencies) 
than the debris ring \citep[cf][]{Kalas2005,Kalas2008,Boley2012}, \textbf{\textit{Fomalhaut b is not an unbound dust cloud.}}

\subsubsection{Constraints on the Orbit of Fomalhaut b Derived from ACS Data}
Comparing the 2004 and 2006 ACS astrometry implies a deprojected space velocity of $v$ $\approx$ 
3.7 $\pm$ 1.4 $km$ s$^{-1}$ or 33\% slower than that derived by \citeauthor{Kalas2008}.  
If Fomalhaut b is a planet sculpting the debris disk, it should be nearly apsidally 
aligned with the disk and have a space velocity of $\sim$ 3.9 km s$^{-1}$ \citep{Chiang2009}.  While 
\citeauthor{Kalas2008}'s velocity is $\sim$ 2-$\sigma$ deviant from apsidal alignment
(5.5$^{+1.1}_{-0.7}$ $km$ s$^{-1}$), ours is consistent with that of an apsidally-aligned object 
within errors.

Next, we compare Fomalhaut b's motion with the debris ring geometry, calculating the minimum distance 
between the companion and the disk midpoint (in angular separation).
We explored a range of PSF subtraction methods for F606W 2004-2006 data sets to identify the 
combination that yields a high SNR image of the disk with minimal levels of self-subtraction.  
Based on these criteria, we use the a PSF subtracted image of the 2006 F606W data obtained under 
the 3\farcs{}0 spot with Vega as the reference star.  

To model the disk and compute the planet-disk separation, we explored two methods to assess 
the robustness of our results.  
\textbf{Method 1} -- We first select a wide annulus enclosing the disk emission, 
mask pixels with intensities roughly less than the 1-$\sigma$ noise floor 
(0.005 cts/s) and
determine a least-squares ellipse fit to the data.
We perform these steps for the 
raw image and a 10-pixel gaussian smoothed image, average the results, and 
report the differences in the results as the fit uncertainties.  
\textbf{Method 2} -- We divided the disk 
into 0.5$^{o}$ annular sections
and
fit the disk center at each section four different ways: 
\begin{itemize} 
\item \textbf{1} A high-order polynomial fit to the 
surface brightness (SB) as a function of distance, taking the maximum value as the disk spine. 
\item \textbf{2} A gaussian fit to the SB with the maximum of the gaussian as the disk spine.
\item \textbf{3} The peak pixel value
\item \textbf{4} The mean distance from all points $>$ 90\% the value of the peak pixel as identifying the disk spine.
\end{itemize}
We then averaged the separately-determined disk radii, calculated x,y positions 
of the disk spine and uncertainties, and derived a least-squares ellipse fit.
As with Method 1, we performed these steps on the raw image (5-pixel binning) 
and a gaussian smoothed image (3-pixel binning).   


Figure \ref{seds} (right panel) depicts our ellipse fitting from Method 1.  Compared to \citet{Kalas2005}, 
we obtain similar values for the disk properties.  From Method 1,
we obtain  137.6 $\pm$ 0.5 AU, 61.6 $\pm$ 0.1 AU, and 156.7 $\pm$ 0.2$^{o}$ for the disk 
major axis, minor axis, and position angle.  From Method 2, we obtain 
140.9 $\pm$ 1.7 AU, 58.0 $\pm$ 0.3 AU, and 156.6 $\pm$ 0.2$^{o}$.
From Method 1, Fomalhaut b appears to move closer to the disk by 0\farcs{}026 
(1\farcs715 to 1\farcs689), whereas 
Method 2 yields an even smaller deviation of 0\farcs002 (1\farcs712 to 1\farcs710).
This change is less than or equal to our quadrature-added astrometric errors for 
either the coordinate ($\sigma_{2006+2004}$ = [0.026, 0.021]) and is not statistically significant.  
Thus, ACS astrometry is consistent with Fomalhaut b being in a ring-nested orbit 
and near apsidal alignment.  However, uncertainties in the astrometry and disk fitting and 
the small number of data points do not yet rule out a ring-crossing orbit
\footnote{\citet{Kalas2010} reported preliminary results indicating that Fomalhaut b 
is at an implausibly wide separation in the 2010 STIS data given their 
ACS astrometry.  \citet{Janson2012} interpret this as evidence that Fomalhaut b is 
in a ring-crossing orbit.  While a detailed analysis of the Fomalhaut STIS 
data is beyond the scope of this paper, our preliminary examination does not cast 
clear doubt on our results.  We recover Fomalhaut b ($m_{STIS, 606W}$ $\approx$ 24.8 $\pm$ 0.2).  
Fomalhaut b's position is significantly 
less discrepant compared to the disk ($\approx$ 0\farcs{}1--0\farcs{}15 $\pm$ 0\farcs{}1) if we adopt the ACS ellipse parameters.
Any uncorrected image distortion, clearly flagged as a problem by \citeauthor{Kalas2010},
will bias image registration by yielding erroneous determinations for where the diffraction spikes intersect (e.g. 
the stellar centroid).  STIS astrometry of Fomalhaut b obtained at multiple epochs 
will better clarify its orbit (Kalas et al. 2012, in prep.).}.  

\section{Discussion}
\citet{Kalas2008}
argue that scattered starlight in a circumplanetary ring system 
contributes significantly to the F606W emission and planet 
thermal radiation dominates the F814W emission.   They find the 0.6 $\mu m$ emission 
variable by 0.7--0.8 mag, explainable by accretion-driven 
$H_{\alpha}$ emission, not dust emission.  In contrast, \citet{Janson2012} argue
that planet thermal emission cannot measurably contribute to the 0.6--0.8 $\mu m$ emission 
else Fomalhaut b (if a planet) should have been detected at 4.5 $\mu m$.  
Based on the IR nondetections, the large implied space velocity from the ACS data, the 
F606W variability, and the difficulty in a circumplanetary ring scattering enough starlight, 
\citeauthor{Janson2012} instead argue that Fomalhaut b is a (semi-)transient dust cloud
physically unassociated with any planet.

We recover \citeauthor{Kalas2008}'s detections at F606W and F814W and 
add a F435W detection, confirming that Fomalhaut b is a real object orbiting the primary.  
We cannot confirm that Fomalhaut b's F606W emission
 is variable at a level larger than our combined photometric 
errors ($\approx$ 0.15 mag), let alone by 0.7--0.8 mags, and 
thus do not find any evidence for accretion-driven emission \textit{nor} for a 
\textit{semi-transient} dust cloud.

Based on the IRAC and (to a lesser extent) $J$-band upper limits, we agree 
that Fomalhaut b's optical emission contains no measurable contribution from a planetary photosphere.
  Thus, in the strictest sense of the word, Fomalhaut b is 
not yet a \textit{directly imaged} planet.
However, a 0.5 $M_{J}$ planet -- favored by \citeauthor{Chiang2009}'s 
simulations -- is too faint to be detectable in \textit{any} existing IR data set.
\textbf{\textit{Thus, the IR non-detections provide no compelling 
evidence against Fomalhaut b being the planet sculpting Fomalhaut's 
debris ring}}.  

Fomalhaut b's ACS positions appear to be consistent with those expected for a ring-nested orbit.  
Moreover, Fomalhaut b's implied space motion is significantly slower and much more consistent with 
that expected for an apsidally aligned object.  Our astrometry then 
weakens the other argument in \citet{Janson2012}'s against Fomalhaut b being a planet sculpting the debris ring.

Finally, Fomalhaut b is unlikely to be an unbound dust cloud since such a cloud 
would shear out into a resolvable clump on very short timescales 
and eventually into a circumstellar ring \citep[see also][]{Kuchner2010}.
Alternatively, a dust cloud gravitationally bound to a planet \textit{can} 
in principle explain the optical SED and persist for much longer timescales.
Though dust confined to an optically-thick circumplanetary ring \citep{Kalas2008} may 
not scatter enough starlight \citep[][]{Janson2012}, a spherical cloud of 
circumplanetary dust originating from ongoing planetesimal collisions \citep{Kennedy2011} is still 
plausible.  

 Fomalhaut b is a rare, possibly unique object in the primary's circumstellar 
environment.  Although the emission identifying it originates from dust, not a planet 
atmosphere, it just happens to be where a massive planet should be in order to 
sculpt the debris ring.  If a planet, it is (as yet) an indirect detection, 
but this is a feature 
shared by the protoplanet LkCa 15 b \citep{Kraus2012}
and radial-velocity/transit/microlensing-detected planets.
Thus, Fomalhaut b is very plausibly (even 'likely') ``a planet 
identified from direct imaging" 
even if images of it, strictly speaking, do not show thermal emission from a directly imaged planet. 

\acknowledgements 
We thank Ray Jayawardhana, Paul Kalas, Mark Wyatt, Christian Thalmann, David Golimowski, Glenn Schneider 
and the anonymous referee for valuable comments.   The STScI help desk staff graciously answered 
detailed queries regarding the HST/ACS data products and flux calibration.

{}

\begin{deluxetable}{lcllccccccc}
\setlength{\tabcolsep}{0pt}
\tablecolumns{4}
\tablecaption{Fomalhaut b Photometry}
\tiny
\tablehead{{UT Date}&{Filter }&{ Processing Method}&{Photometry (Vega)}}
\startdata
20090816 & J & 1 & $>$ 22.22 (5-$\sigma$)\\
20060714-20& F435W & 3 & 25.22 $\pm$ 0.18 & -\\
"& F606W & 2,4 & 24.97 $\pm$ 0.09\\
"& F814W & 3 & 24.91 $\pm$ 0.20 & \\
20040926 & F606W & 4 & 24.92 $\pm$ 0.10\\
 \enddata
\tablecomments{Processing Methods: 1) LOCI, 2) Reference PSF Subtraction, 3) Roll Subtraction, 4) and A-LOCI.}
\label{hr8799prop}
\end{deluxetable}


\end{document}